# Design of a Fractional Order PID Controller Using Particle Swarm Optimization Technique

#Deepyaman Maiti, Sagnik Biswas, Amit Konar
*Department of Electronics and Telecommunication Engineering, Jadavpur University*
*Kolkata - 700 032*
*deepyamanmaiti@gmail.com, sagnik_agp@rediffmail.com, konaramit@yahoo.co.in*

**Abstract**

*Particle Swarm Optimization technique offers optimal or sub-optimal solution to multidimensional rough objective functions. In this paper, this optimization technique is used for designing fractional order PID controllers that give better performance than their integer order counterparts. Controller synthesis is based on required peak overshoot and rise time specifications. The characteristic equation is minimized to obtain an optimum set of controller parameters. Results show that this design method can effectively tune the parameters of the fractional order controller.*

## 1. INTRODUCTION

Proportional-Integral-Derivative (PID) controllers are widely being used in industries for process control applications. The merit of using PID controllers lie in its simplicity of design and good performance including low percentage overshoot and small settling time for slow industrial processes. The performance of PID controllers can be further improved by appropriate settings of fractional-I and fractional-D actions. This paper attempts to study the behavior of fractional PID controllers over integer order PID controllers.

In a fractional PID controller, the I- and D-actions being fractional have wider scope of design. Naturally, besides setting the proportional, derivative and integral constants $K_p$, $T_d$ and $T_i$ respectively, we have two more parameters: the power of s in integral and derivative actions- $\lambda$ and $\delta$ respectively. Finding $[K_p, T_d, T_i, \lambda, \delta]$ as an optimal solution to a given process thus calls for optimization on the five-dimensional space. Classical optimization techniques cannot be used here because of the roughness of the objective function surface. We, therefore, use a derivative-free optimization, guided by the collective behavior of social swarm and determine optimal settings of $K_p$, $T_d$, $T_i$, $\lambda$ and $\delta$.

The performance of the optimal fractional PID controller is better than its integer counterpart. Thus the proposed design will find extensive applications in real industrial processes. Traces of work on fractional PID are available in the current literature [1] – [9] on control engineering. A frequency domain approach based on the expected crossover frequency and phase margin is mentioned in [2]. A method based on pole distribution of the characteristic equation in the complex plane was proposed in [5]. A state-space design method based on feedback poles placement can be viewed in [6]. The fractional controller can also be designed by cascading a proper fractional unit to an integer-order controller.

Our design focuses on positioning closed loop dominant poles, and the constraints thus obtained on the characteristic equation are optimally satisfied by particle swarm optimization algorithm. The work is thus original and may open up new avenues for the next generation fractional order controller design.

It is necessary to understand the theory of fractional calculus in order to realize the significance of fractional order integration and derivation. Fractional calculus is the branch of calculus that generalizes the derivative or integral of a function to non-integer order, allowing calculations such as deriving a function to 1/2 order. Since $s^\alpha$ indicates deriving to the order $\alpha$, knowledge in the subject of fractional calculus is essential to design fractional order controllers.

Of the several definitions of fractional derivatives, the Grunwald-Letnikov and Riemann-Liouville definitions are the most used. These definitions are required for the realization of discrete control algorithms.

## 2. THE INTEGER AND FRACTIONAL ORDER PID CONTROLLERS

The integer order PID controller has the following transfer function: $K_p + T_i s^{-1} + T_d s$.

Here, the orders of integration and derivation are both unity.

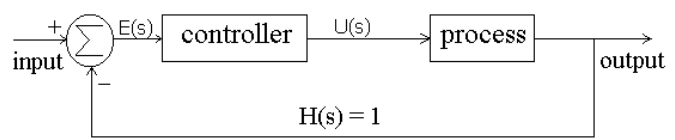

Fig. 1: Generic Closed Loop System

The real objects or processes that we want to control are generally fractional (for example, the voltage-current relation of a semi-infinite lossy RC line). However, for many of them the fractionality is very low.

In general, the integer-order approximation of the fractional systems can cause significant differences between mathematical model and real system. The main reason for using integer-order models was the absence of solution methods for fractional-order differential equations.

PID controllers belong to dominating industrial controllers and therefore are objects of steady effort for improvements of their quality and robustness. One of the possibilities to improve PID controllers is to use fractional-order controllers with non-integer derivation and integration parts.

A fractional PID controller therefore has the transfer function:

$$K_p + T_i s^{-\lambda} + T_d s^{\delta}.$$

The orders of integration and differentiation are respectively $\lambda$ and $\delta$ (both positive real numbers, not necessarily integers). Taking $\lambda = 1$ and $\delta = 1$, we will have an integer order PID controller. So we see that the integer order PID controller has three parameters, while the fractional order PID controller has five.

The fractional order PID controller generalizes the integer order PID controller and expands it from point to plane. This expansion adds more flexibility to controller design and we can control our real world processes more accurately.

We will design both integer order and fractional order PID controllers using the particle swarm optimization (PSO) algorithm and display the advantages the fractional order controllers provide us over the integer order controllers.

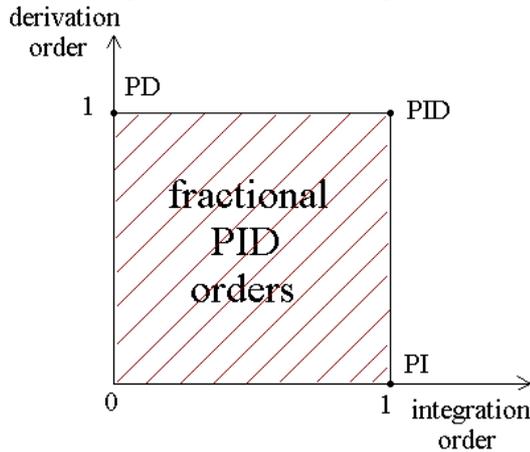

Fig. 2. Expanding from Point to Plane

### 3. STANDARD PSO ALGORITHM

The PSO algorithm [10] - [11] attempts to mimic the natural process of group communication of individual knowledge, which occurs when a social swarm elements flock, migrate, forage, etc. in order to achieve some optimum property such as configuration or location.

The 'swarm' is initialized with a population of random solutions. Each particle in the swarm is a different possible set of the unknown parameters to be optimized. Representing a point in the solution space, each particle adjusts its flying toward a potential area according to its own flying experience and shares social information among particles. The goal is to efficiently search the solution space by swarming the particles toward the best fitting solution encountered in previous iterations with the intent of encountering better solutions through the course of the process and eventually converging on a single minimum error solution.

Let the swarm consist of N particles moving around in a D-dimensional search space. Each particle is initialized with a random position and a random velocity. Each particle modifies its flying based on its own and companions' experience at every iteration. The $i^{th}$ particle is denoted by $X_i$, where $X_i = (x_{i1}, x_{i2}, \ldots, x_{iD})$. Its best previous solution (pbest) is represented as $P_i = (p_{i1}, p_{i2}, \ldots, p_{iD})$. Current velocity (position changing rate) is described by $V_i$, where $V_i = (v_{i1}, v_{i2}, \ldots, v_{iD})$. Finally, the best solution achieved so far by the whole swarm (gbest) is represented as $P_g = (p_{g1}, p_{g2}, \ldots, p_{gD})$.

At each time step, each particle moves towards pbest and gbest locations. The fitness function evaluates the performance of particles to determine whether the best fitting solution is achieved. The particles are manipulated according to the following equations:

$$v_{id}(t+1) = \omega v_{id}(t) + c_1.\varphi_1.(p_{id}(t) - x_{id}(t)) + c_2.\varphi_2.(p_{gd}(t) - x_{id}(t)) \quad (1)$$

$$x_{id}(t+1) = x_{id}(t) + v_{id}(t+1). \quad (2)$$

(The equations are presented for the $d^{th}$ dimension of the position and velocity of the $i^{th}$ particle.)

Here, $c_1$ and $c_2$ are two positive constants, called cognitive learning rate and social learning rate respectively, $\varphi_1$ and $\varphi_2$ are two random functions in the range $[0,1]$, $\omega$ is the inertia factor which balances the global wide-range exploitation and the local nearby exploration abilities of the swarm.

### 4. APPLICATION OF THE PSO ALGORITHM TO THE PROBLEM OF DESIGNING PID CONTROLLERS

Our approach is based on the root locus method (dominant roots method) of designing integral PID controllers.

As in the traditional root locus method, based on our requirements of peak overshoot $M_p$ and rise time $t_{rise}$ (or requirements of stability and damping levels), we find out the damping ratio $\zeta$ and the undamped natural frequency $\omega_0$. Using the values of $\zeta$ and $\omega_0$ we then find out the positions of the dominant poles of the closed loop system,

$$p_{1,2} = -\zeta\omega_0 \pm j\omega_0\sqrt{1-\zeta^2}. \quad (3)$$

Let the closed loop transfer function of the system is:

$$\frac{G(s)}{1+G(s)H(s)}$$

Here $G(s) = G_c(s).G_p(s)$ where $G_c(s)$ is the transfer function of the controller to be designed. $G_c(s)$ is of the form

$$G_c(s) = K_p + T_i s^{-\lambda} + T_d s^{\delta}. \quad (4)$$

$G_p(s)$ is the transfer function of the process we want to control.

If the required closed loop dominant poles are located at $s_{1,2} = p_{1,2} = -x + jy, -x - jy$, then at $s = p_1 = -x + jy$, we must have

$$1 + G(p_1).H(p_1) = 0. \quad (5)$$

From (5), we get:

$$1 + (K_p + T_i s^{-\lambda} + T_d s^{\delta}).G_p(p_1).H(p_1) = 0. \quad (6)$$

Assuming $H(s) = 1$, and $G_p(s)$ being known, (6) can be arranged as:

$$1+[K_p+T_i(-x+jy)^{-\lambda}+T_d(-x+jy)^{\delta}]G_p(-x+jy)= 0. \quad (7)$$

In this complex equation (7) we have five unknowns, namely $\{K_p, T_i, T_d, \lambda, \delta\}$. There are an infinite number of solution sets for $s = p_1 = -x + jy$. So the equation cannot be unambiguously solved.

At this juncture, the PSO algorithm helps us the find the optimal solution set to the complex equation.

Let:
R=real part of the complex expression,
I=imaginary part of the complex expression,
P=phase ( $= \tan^{-1}(I/R)$ ).
We define $f = |R| + |I| + |P|$ and minimize 'f' using the PSO technique. Our goal is to find out the optimum solution set $\{K_p, T_i, T_d, \lambda, \delta\}$ for which f=0.

The solution space is five-dimensional, the five dimensions being $K_p$, $T_i$, $T_d$, $\lambda$ and $\delta$. So each particle has five-dimensional position and velocity vectors. The personal and global bests are also five-dimensional. The limits on the position vectors of the particles (i.e. the controller parameters) are set by us as follows.

As a practical assumption, we allow $K_p$ to vary between 1 and 1000, $T_i$ and $T_d$ between 1 and 500, $\lambda$ and $\delta$ between 0 and 2. Initializations of the five variables are also done in the above-mentioned ranges.

We also set the inertia factor $\omega=0.729$ and $c_1=c_2=1.494$.

After running the PSO algorithm, we obtain the position vector of the best particle i.e. the optimized values of the five controller parameters.

For tuning an integer order PID controller, we simply set $\lambda = \delta = 1$, so that the solution space becomes three-dimensional. The search ranges for the other three variables and the values of $\omega$, $c_1$ and $c_2$ are kept same as before.

After running the PSO algorithm, we obtain the position vector of the best particle i.e. the optimized values of the three controller parameters.

## 5. ILLUSTRATIONS

*A. Example 1*

The process (control objective) has the transfer function

$$\frac{1}{0.8s^{2.2} + 0.5s^{0.9} +1}$$

We want to design a controller such that the closed loop system has a maximum peak overshoot $M_p = 10\%$ and $t_{rise} = 0.3$ seconds.

This translates to $\zeta = 0.65$ and $\omega_0 = 2.2$ s$^{-1}$. The dominant poles for the closed loop controlled system should lie at $(-1.43 + j1.67)$ and $(-1.43 - j1.67)$.

For $p_1 = (-1.43 + j1.67)$, the characteristic equation is:

$$1+\frac{K_p + T_i(-1.43+j1.67)^{-\lambda} + T_d(-1.43+j1.67)^{\delta}}{0.8(-1.43+j1.67)^{2.2} + 0.5(-1.43+j1.67)^{0.9} + 1} = 0 \quad (8)$$

After separating the real and imaginary parts we have:

$$R = (K_p+1)+\frac{T_i}{2.2^{\lambda}}\cos(130.57\lambda)° + T_d 2.2^{\delta}\cos(130.57\delta)° + 0.875 \quad (9)$$

$$I = -\frac{T_i}{2.2^{\lambda}}\sin(130.57\lambda)° + T_d 2.2^{\delta}\sin(130.57\delta)° - 3.428 \quad (10)$$

Also, $P = \tan^{-1}(I/R)$. (11)

We minimize $f = |R| + |I| + |P|$ using PSO technique using the following limits:

$1 \leq K_p \leq 1000$, $1 \leq T_i, T_d \leq 500$, $0 \leq \lambda, \delta \leq 2$, $\omega=0.729$ and $c_1=c_2=1.494$.

The optimized parameters for the fractional order PID controller are:
$K_p=442.68$, $T_i=324.03$, $T_d=115.27$, $\lambda=1.5$, $\delta=1.41$.

The transfer function for the fractional order PID controller is: $442.38 + 324.03s^{-1.5} + 115.27s^{1.41}$.

If we set $\lambda = 1$ and $\delta = 1$ before running the PSO algorithm, we obtain the three optimized parameters for the integer order PID controller.

The optimized parameters for the integer order PID controller are:
$K_p = 214.84$, $T_i = 361.57$, $T_d = 76.76$.

The transfer function for the integer order PID controller is: $214.84 + 361.57s^{-1} + 76.76s$.

Finally we plot the time responses for unit step input in Fig. 3 for:
 uncontrolled system open loop response,
 system controlled by integral PID controller,
 system controlled by fractional PID controller.

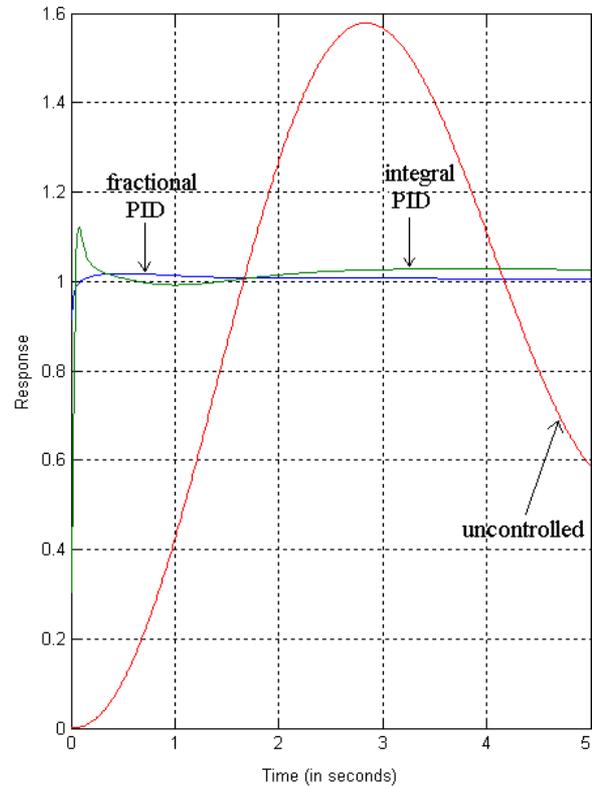

Fig. 3. Closed Loop Unit Step Response for Example 1

For the integer order PID controlled process, the maximum peak overshoot is 12.5% and the rise time is 0.05 seconds.

For the fractional order PID controlled process, the maximum peak overshoot is less than 2% and the rise time is 0.1 seconds.

### B. Example 2

The process has the transfer function:

$$\frac{400}{s^2 + 50s}$$

We want to design a controller such that the closed loop system has a maximum peak overshoot $M_p = 10\%$ and $t_{rise} = 0.3$ seconds.

Proceeding as before:

The optimized parameters for the fractional order PID controller are:

$K_p = 32.01$, $T_i = 10.14$, $T_d = 9.71$, $\lambda = 1.19$, $\delta = 1.36$.

The transfer function for the fractional order PID controller is:

$$32.01 + 10.14s^{-1.19} + 9.71s^{1.36}$$

Setting $\lambda = 1$ and $\delta = 1$ before running the PSO algorithm, we obtain the optimized parameters for the integer order PID controller as:

$K_p = 3.2$, $T_i = 5.41$, $T_d = 1$.

The transfer function for the integer order PID controller is:

$$3.2 + 5.41s^{-1} + s.$$

Finally we plot the time responses for unit step input in Fig. 4 for:

system controlled by integral PID controller,
system controlled by fractional PID controller.

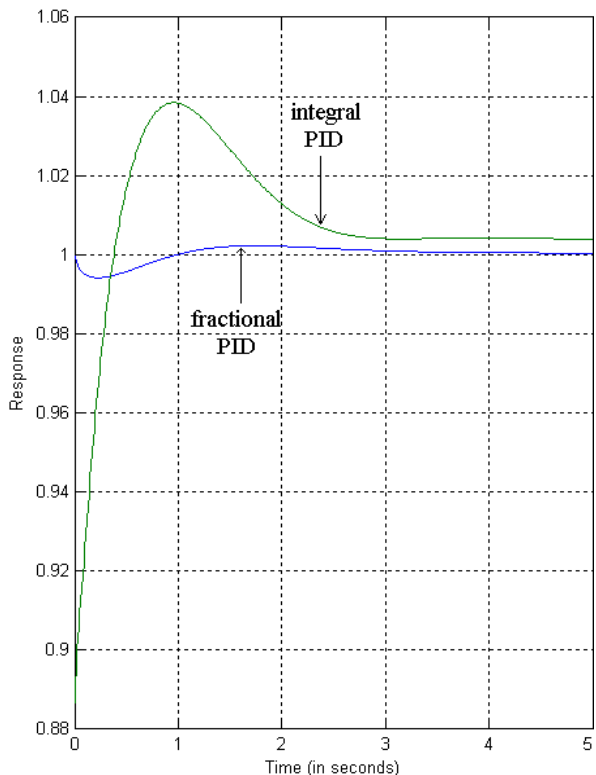

Fig. 4. Closed Loop Unit Step Response for Example 2

For the integer order PID controlled process, the maximum peak overshoot is 4% and the rise time is 0.4 seconds.

For the fractional order PID controlled process, the maximum peak overshoot is about 0.5% and the rise time is almost nil.

## 6. RESULTS

Using fractional order PID controllers, we have significantly reduced percentage overshoot and rise and settling times (compared to integral PID controllers).

The controllers were designed using the dominant pole in the second quadrant (i.e. $-x + jy$). Similar results were obtained when the third quadrant dominant pole $-x - jy$ was used.

After running the PSO algorithm a substantial number of times, it was found that almost all the particles had fitness zero or very close to zero.

It is noted that for the given common performance criteria on $M_p$ and $t_{rise}$, the fractional order controller achieves better results than its integer counterpart. The proposed scheme of fractional PID controller design will thus find extensive commercial application in the next generation controller design.